# Quantum Decision Theory

**Hervé Zwirn**
Centre Borelli (ENS Paris Saclay) & IHPST (UMR 8590 CNRS / Paris 1 / ENS Ulm)
herve.zwirn@gmail.com

November 2024

**Summary**
In this article, we propose to use the formalism of quantum mechanics to describe and explain the so-called "abnormal" behaviour of agents in certain decision or choice contexts. The basic idea is to postulate that the preferences of these agents are indeterminate (in the quantum sense of the term) before the choice is made or the decision is taken. An agent's state before the decision is represented by a superposition of potential preferences. The decision is assimilated to a measure of the agent's state and leads to a projection of the state onto one of the particular preferences. We therefore consider that uncertainty about preferences is not linked to incomplete information but to essential indeterminacy. We explore the consequences of these hypotheses on the usual concepts of decision theory and apply the formalism to the problem of the so-called "*framing*" effect.

## 1  Introduction

It is now well known that people involved in making choices or decisions sometimes behave in apparently irrational ways, which we refer to as behavioural anomalies [1]. These anomalies can manifest themselves in different ways. Faced with the same possibilities, an agent may, for example, make a different choice depending on the way in which the choice is presented to him, or depending on the context in which he is placed, even though the different contexts appear to be equivalent in terms of this choice. It is also possible that, having to make several successive independent choices, the agent will end up with results that depend on the order in which these choices are made. Conventional approaches find it difficult to account for these anomalies without resorting to arguments that often seem ad hoc.

The traditional Bayesian approach to modelling incomplete information, suggested by Harsanyi [2], consists of adopting an a priori probability distribution on the types of agents (a type being supposed to represent all the relevant information related to the agent in the decision situation under consideration), drawing lots from the types and informing each agent of her own type. The result is that uncertainty about an agent's type mainly reflects the incomplete information that other agents have about her. This is because a type is perfectly determined and represents the complete and well-defined set of characteristics of an agent. Each agent knows her own type, but only has a probability distribution for the types of the other agents. It is on this point that we depart from the traditional approach in that we assume that, in addition to a lack of information, uncertainty about an agent's type may arise from the fact that it is not completely determined before the agent has made her choice or decision. The agent's state is then a superposition of different types in the sense that quantum formalism allows. It is then only at the moment of decision, which we identify with the equivalent of a measurement, that the preference is determined. This idea is in line with what Tversky and Simonson [3] suggest, according to which:

*"There is a growing body of evidence to support an alternative view that preferences are often constructed - not just revealed - at the point of choice. These constructions depend on the framework in which the problem is posed, the method of obtaining the results and the context of the choice".*

This point of view seems to be fairly consistent with the observation that agents (even a priori highly rational ones) can behave differently in equivalent situations that differ only in apparently irrelevant factors (such as the environment or prior events unrelated to the situation under consideration).

## *2* The formal framework

In this section, we present the formalism we will be using and its interpretation [4]. We borrow from quantum mechanics some of the tools that have been developed to model the atomic world[1]. Quantum formalism is largely based on the Hilbert space model, which is the natural structure for expressing the state of a system in the form of a vector. This model is supplemented in physics by a dynamic equation describing how the state evolves over time, the Schrödinger equation, but we will not have an equivalent in the framework proposed here. We will therefore only use the Hilbert space formalism, to which we will add a few additional rules, such as the one governing measurement. It is also known that this formalism can be considered as a generalisation of the probability calculus when we want to introduce contextual dependencies. The fact that it is suitable for describing the observations mentioned above is therefore not so surprising, but we will not have the opportunity to go into this remark in more detail, as it would require technical developments that have no place here (see [5] for example).

### 2.1 The notion of state and superposition

What we are trying to describe is the choice behaviour of an agent in a decision situation, which we interpret as a revelation of her preferences. In this article, we will restrict ourselves to situations where a choice has to be made in a non-repetitive way and without taking into account any notion of strategy. We therefore exclude game situations where a player has to respond iteratively to a choice made by the opposing player. These situations are more complex and require further development, which is the subject of ongoing work. Examples of situations that fall within the scope of this article are:
- choose between purchasing a M1 or M2 or M3 brand computer
- choose whether or not to invest in a project
- choose between a certain win of €100 or a bet giving €250 with a probability of 0.5 and €0 with a probability of 0.5
- prefer to eat a banana, an apple or a pear
- choose to cooperate or to denounce in the prisoner's dilemma

In the rest of the article, the examples we give to illustrate our point will always be elementary. Of course, this in no way affects the possibility of using the formalism in more sophisticated cases. An agent is represented by a state that encompasses everything that can be known about her expected behaviour in the situation under consideration. In the simplest classical case, the state could be directly the designation of the choice she will make according to the different possibilities. In this case, the state represents the agent's preference for a particular choice. For example, if the decision situation consists of choosing between a banana, an apple or a pear, the agent's state could be "pear". This would mean that, faced with the choice in question, the agent would choose the pear with certainty. A more interesting situation is that modelled by Bayesian formalism, in which each possibility has a probability of being chosen. The state could then be "banana with probability 0.3, pear with probability 0.2 and apple with probability 0.5". But it is also possible for the state to contain an order of preference, such as "pear, banana, apple", meaning that the agent will prefer a pear to any other fruit, but will choose a banana over an apple.

---

[1] We assume the reader to have a basic knowledge of the quantum formalism.

By analogy with the case of quantum mechanics, we will mathematically represent the agent's state as a vector in a Hilbert space $\mathcal{H}$ and denote it $|\psi\rangle$. The link between the Hilbert space $\mathcal{H}$ and the decision situation will be explained below. At this stage, what exactly the state $|\psi\rangle$ represents is left open. Knowledge of the agent's state should, in principle, allow predictions to be made about what she will do when faced with a certain decision situation. We will see that the interpretation that can be given can vary according to the context and the modelling that we wish to do. According to the principle of superposition, if $|\psi_1\rangle$ and $|\psi_2\rangle$ are two possible states of the agent, any linear combination $\lambda_1|\psi_1\rangle + \lambda_2|\psi_2\rangle$ avec $\lambda_1^2 + \lambda_2^2 = 1$ is also a possible state. This already means that, even in the case of the simplest interpretation mentioned above, i.e. where the state directly represents the choice that the agent will make, it is possible to obtain, by linear combination of such states, states (known as superposed states) that can no longer be interpreted classically. These are superpositions, for example, of a state representing the sure choice of an apple with a state representing the sure choice of a banana. Such states, which are impossible to conceive of in a classical framework and which we must not attempt to understand, are the very essence of the difference between the usual formalisms and the one we are proposing.

## 2.2 The notion of observable and measurement

A measurement is an operation carried out on the system which produces a result. Typically, in physics, a measurement is carried out using a device that is supposed to determine the value of a physical quantity, such as a position, a momentum or a spin. The characteristic of a measurement operation is that if it is repeated immediately afterwards, without a different measurement having been made, it will give the same result. Every physical quantity in the system has an associated observable whose eigenvalues give the only possible results that can be obtained when that quantity is measured. After a measurement with one of the eigenvalues as the result, the state vector of the system is projected onto the eigensubspace associated with this eigenvalue. When the eigenvalue is not degenerate, the state vector therefore becomes equal to the eigenvector associated with the eigenvalue obtained during the measurement. We will use this mechanism to model a situation in which an agent is faced with a choice between several alternatives, which we will call a "decision situation".

## 2.3 Single decision situation

A decision situation is defined by the set of alternatives from which the agent must choose. It is the situation that will determine the associated Hilbert space and its dimensionality. The corresponding space will be chosen so that its dimension is at least equal to the number of possible choices (greater, if we want to use degenerate eigenvalues as we will see below). We will treat such a situation as a measurement by identifying the choice made with the value produced by the measurement. The process of choice is therefore similar to the measurement of a quantity. An observable A will be associated with each decision situation. If n different choices are proposed, the eigenvectors of A will be conventionally noted $|1\rangle$, ..., $|n\rangle$ and will be associated with eigenvalues 1,..., n with the convention that obtaining eigenvalue j corresponds to having made the choice j in the list of possible choices. Since in this hypothesis none of the eigenvalues is degenerated, {|1⟩, …, |n⟩} is the only orthonormal basis of $\mathcal{H}$ formed by eigenvectors of A. The agent's state can therefore be written in this basis:

$$|\psi\rangle = \sum_{k=1}^{n} \lambda_k |k\rangle \text{ with } \sum_{k=1}^{n} |\lambda_k|^2 = 1$$

According to the reduction principle, the probability that the agent in state |ψ⟩ chooses alternative *i* (i.e. obtains eigenvalue *i*) is: $|\langle i|\psi\rangle|^2 = |\lambda_i|^2$

Immediately after the measurement, the agent's state is projected onto the eigenstate associated with eigenvalue *i*, i.e. $|i\rangle$. If the same decision situation is repeated and the agent is asked to decide again, she will make the same choice with certainty by opting for alternative *i*. In this simple case, our formalism is equivalent to the usual probabilistic formalism in which probabilities are assigned to the agent's various possible choices. The predictions obtained from the state $|\psi\rangle = \sum_{k=1}^{n} \lambda_k |k\rangle$ are identical to those that would be obtained from a classical state described as the fact that each possible choice *k* has probability $|\lambda_k|^2$ to be obtained.

## 2.4 Multiple decisions situations

*Commuting decisions situations*

Let's suppose that the agent is faced with two decision situations (for example, the choice between an apple and a banana on the one hand, and the choice between spending a holiday at the seaside or in the mountains on the other). Let A and B be the observables associated respectively with each of the situations. Let us first assume that the two situations offer the same number n of alternatives, which will allow us to assume that the eigenvalues are not degenerate. If A and B commute, there exists a basis of the Hilbert space $\mathcal{H}$ formed by eigenvectors common to A and B. Let us note $|i\rangle$ these base vectors. Then we have: $A|i\rangle = i_A |i\rangle$ et $B|i\rangle = i_B |i\rangle$ avec $i_A$ et $i_B \in \{1, ..., n\}$

But given our convention that obtaining the eigenvalue *i* when measuring A means obtaining the choice of rank *i* in the list of possible choices of A, we can always order the list of choices of A and B in such a way that $A|i\rangle = B|i\rangle = i|i\rangle$

This means that A=B. Any vector in Hilbert space can be written on this basis and the agent state will be: $|\psi\rangle = \sum_{i=1}^{n} \lambda_i |i\rangle$ with $\sum_{i=1}^{n} |\lambda_i|^2 = 1$

If we measure A first[2], the probability of obtaining the eigenvalue *i* is : $p_A(i) = |\lambda_i|^2$

We will note this choice *i(A)* to avoid any ambiguity. So *i(A)* is the choice of rank *i* of the decision situation A while *i(B)* is the choice of the same rank *i* for the decision situation B. These two choices are associated with the same eigenvalue.

---

[2] By abuse of language we will say: "measuring A" instead of: "making the choice corresponding to the decision situation with which observable A is associated".

The probability $p_A(i)$ of obtaining the eigenvalue $i$ when measuring A first is the same as the probability $p_B(i)$ to obtain the eigenvalue $i$, and therefore the choice $i(B)$, if we measure B first.

Suppose we have measured A and obtained the choice $i(A)$. The agent's state vector after the measurement will be $|i\rangle$ and the probability of obtaining the choice $j(B)$ in a subsequent measurement of B will be $|\langle j|i\rangle|^2 = \delta(j,i)$ (which is 1 if $j=i$ and 0 otherwise). So we can see that in this very simple case, there is a total correlation between the choices in situation A and those in situation B. As the choice of rank $i$ is made for A, we are sure to obtain the choice of rank $i$ for B (and vice versa), which follows directly from the fact that A=B. This trivial case is not very interesting and we must generalise the way we model the sates.

In order to relax this constraint, it is necessary to use degenerate eigenvalues, which will also allow us to deal with the case where the number of choices in each of the situations is not the same. In this case we will note:

$$A|i\rangle = i_A|i\rangle \text{ and } B|i\rangle = i_B|i\rangle$$

so that $i_A$ is the eigenvalue of A associated with the eigenvector $|i\rangle$ and knowing that an eigenvalue of A will be degenerate if, for at least one pair $(i, j)$ with $i \neq j$, we have $i_A = j_A$ (idem for B). The number of eigenvectors of A is then greater than the number of possible alternatives in the decision situation associated with A [3].

In this case, the probability of obtaining the choice $i(A)$, if we measure A first, becomes:

$$p_A(i) = \sum_{j:j_A=i} |\lambda_j|^2$$

If we measure B first, the probability of obtaining the choice $j(B)$ is:

$$p_B(j) = \sum_{k:k_B=j} |\lambda_k|^2$$

After obtaining $j(B)$, the agent's state becomes:

$$|\psi_j\rangle = \frac{1}{\sqrt{\sum_{k:k_B=j}|\lambda_k|^2}} \sum_{k:k_B=j} \lambda_k |k\rangle$$

If we then measure A for an agent in this state, the probability of obtaining $i(A)$ is [4]:

$$p_{AB}(i|j) = \frac{1}{\sum_{k:k_B=j}|\lambda_k|^2} \sum_{k:k_B=j \text{ et } k_A=i} |\lambda_k|^2$$

so the probability of obtaining $i$ when A is measured after B is measured is:

---

[3] The non-degenerate case corresponds to a situation with n alternatives where $i_A = i \ \forall i \in \{1,...,n\}$.

[4] We note $p_{AB}(i|j)$ the conditional probability of obtaining $i$ by measuring A when $j$ has been obtained by first measuring B.

$$p_{AB}(i) = \sum_j p_B(j) p_{AB}(i|j) = \sum_j \left[ \sum_{k:k_B=j} |\lambda_k|^2 \frac{1}{\sum_{k:k_B=j} |\lambda_k|^2} \sum_{k:k_B=j \text{ and } k_A=i} |\lambda_k|^2 \right]$$

$$= \sum_j \sum_{k:k_B=j \text{ and } k_A=i} |\lambda_k|^2 = \sum_{k:k_A=i} |\lambda_k|^2 = p_A(i)$$

We can therefore see that $p_{AB}(i) = p_A(i)$ which means that measuring B before A changes nothing in the measurement of A (idem for B). When the observables associated with two decision situations commute, we can measure one and the other independently. The measurement of one has no influence on the measurement of the other. We can also measure both and determine the joint probability $p_{AB}(i \wedge j) = \sum_{k:k_A=i \text{ and } k_B=j} |\lambda_k|^2$. This means that the event of measuring $i$ on A and $j$ on B is well defined. We can therefore merge the two situations and define a probability space over the pairs $(i, j)$. In addition, the conditional probability formula $p_{AB}(i \wedge j) = p_A(i) p_B(j|i)$ applies. The introduction of degenerate eigenvalues therefore makes it possible to free ourselves from the constraint of total correlation (and even identity) that we had initially observed and to deal with the general framework in which all combinations of choices between A and B are possible.

Our formalism in the case of commuting decision situations therefore reproduces the results of the classical Bayesian formalism in which we are given a probability distribution on the joint choices of situations. This necessary observation shows the consistency of our proposed extension with respect to its classical limit.

*Non-commuting decision situations*

It is in the case of non-commuting decision situations that the quantum formalism provides new predictions compared with the classical formalism. As we shall see, the differences arise from interference terms in the calculation of probabilities. Suppose we have two decision situations A and B associated with observables A and B that do not commute, and that these observables have the same number n of non-degenerate eigenvalues[5] (which we can therefore denote 1, 2, ..., n). Unlike the previous case, there is no longer a basis of eigenvectors common to A and B. Let us therefore note $\{|1_A\rangle, |2_A\rangle, ..., |n_A\rangle\}$ the basis of eigenvectors of A, the eigenvector $|i_A\rangle$ being associated with the eigenvalue $i$ (idem for B). The agent's initial state can be written on each of these bases:

$$|\psi\rangle = \sum_{i=1}^n \lambda_i |i_A\rangle = \sum_{j=1}^n \nu_j |j_B\rangle \text{ with } \sum_{i=1}^n |\lambda_i|^2 = 1 \text{ et } \sum_{j=1}^n |\nu_j|^2 = 1$$

The eigenvectors of B can be written in the basis of the eigenvectors of A:

$$|j_B\rangle = \sum_{i=1}^n \mu_{ij} |i_A\rangle$$

So:

---

[5] In this case, the introduction of degenerate eigenvalues, which is necessary when the number of possible choices is not the same for the two situations, does not change anything essential in the results we are going to show. On the other hand, it does complicate the notation to be used.

$$|\psi\rangle = \sum_{j=1}^{n} v_j |j_B\rangle = \sum_{j=1}^{n} \sum_{i=1}^{n} v_j \mu_{ij} |i_A\rangle$$

If the agent is first in decision situation A, he will choose choice $i$ with probability:

$$p_A(i) = \left| \sum_{j=1}^{n} v_j \mu_{ij} \right|^2$$

If the agent is first in decision situation B, he will choose choice $j$ with probability $|v_j|^2$ and her state will be projected onto the vector $|j_B\rangle$. The probability that, faced then with decision situation A, he will choose choice $i$ is $|\mu_{ij}|^2$. It follows that the probability that the agent having first chosen $j$ in decision situation B, will then choose choice $i$ in situation A is:

$$p_{AB}(i) = \sum_{j=1}^{n} |v_j|^2 |\mu_{ij}|^2 = \sum_{j=1}^{n} |v_j \mu_{ij}|^2$$

which is generally different from $p_A(i)$ which contains cross terms (interference terms):

$$p_A(i) = \left| \sum_{j=1}^{n} v_j \mu_{ij} \right|^2 = p_{AB}(i) + \sum_{j \neq j'} v_{j'}^* \mu_{ij'}^* v_j \mu_{ij'}$$

As a result, the agent's choice in situation A will generally be different depending on whether or not she has previously been confronted with situation B. Furthermore, in the case of two observables which do not commute, we know that it is not possible to consider that the results are simultaneously defined for both (in the sense that a measurement of one or the other will always give the same result). It also follows that the conditional probability formula:

$$p_{AB}(i \wedge j) = p_A(i) p_B(j | i)$$

no longer applies, as can be easily verified since:

$$\left| \sum_{j=1}^{n} v_j \mu_{ij} \right|^2 = p_A(i) \neq \sum_{j=1}^{n} p_B(j) p_A(i | j) = \sum_{j=1}^{n} |v_j \mu_{ij}|^2$$

## *3 Two examples of application*

Let's now look at two examples where the formalism presented can provide a model to explain a surprising result in a natural way. The first example is a fictitious experiment that we propose in order to test our hypotheses. If the possibility for an agent to be in a superposed state of mind (representing a true indeterminacy of its preferences) is real, then an experiment of the type presented below should give a result that does not conform to what classical intuition tells us. Doing a real experiment in a mode similar to the very schematic one we are proposing, and obtaining such a result, would be an appreciable clue to confirm our model. The second example is taken from the literature and concerns the famous *framing* problem, for which we propose a new explanation.

## 3.1 Superposed states of mind

As we shall see, this experiment is a transposition of the double-slit experiment [6] to the context of the decision.

*The experimental set-up*

Consider two identical populations of agents I and II. Suppose that each member of each population is invited to play the prisoner's dilemma[6] against a hidden player. Before playing, the agents in population I must individually answer a question with YES or NO. The detail of the question is of little importance for the purposes of the argument except that the question relates to a typical characteristic in connection with the game. For example, the question could be interpreted as revealing whether the agent is altruistic or selfish. The agents in Population II, on the other hand, play directly without having to answer the question. Let's start with population I. Suppose we have a proportion $\alpha$ of YES and $(1-\alpha)$ of NO answers to the question. The agents then play the prisoner's dilemma. Suppose we have a proportion $\beta$ of agents who answered YES who cooperate (a proportion $(1-\beta)$ denounces). Similarly, suppose we have a proportion of $\gamma$ of agents who answered NO who cooperate (a proportion $(1-\gamma)$ denounces). The proportion of agents in population I who cooperate will therefore be:
$$P_I(coop) = \alpha\beta + (1-\alpha)\gamma$$

Let us now consider population II for which the agents have played directly. We find a proportion of agents who cooperate equal to PII(coop). We would of course expect $P_{II}(coop) = P_I(coop)$. There are several reasons for this. The simplest is that the two populations are identical and therefore must produce the same results unless an event has differentiated them before their agents play. However, the simple fact of answering a question about their character can neither have transformed the agents, nor prompted them to play differently. More detailed reasoning leads to the same conclusion. If we denote $p_{II}(A)$ the proportion of altruistic agents and $p_{II}(E)$ the proportion of selfish agents in population II, we can write according to the law of conditional probabilities:
$$P_{II}(coop) = p_{II}(A)p(coop|A) + p_{II}(E)p(coop|E)$$

But population II is identical to population I. Consequently, even if it has not been measured, it is natural to assume that the proportion of altruistic (respectively egoistic) agents in population II is identical to that in population I for which the measurement has been made (an agent is altruistic or egoistic regardless of whether or not he has been asked the question). So:
$$p_{II}(A) = \alpha; \quad p_{II}(E) = (1-\alpha);$$

Similarly, there is no reason to suppose that the proportion of altruistic agents who cooperate (or denounce) is different in the two populations. It therefore follows that:
$$p(coop|A) = \beta; \quad p(coop|E) = \gamma$$
So:
$$P_{II}(coop) = p_{II}(A)p(coop|A) + p_{II}(E)p(coop|E) = \alpha\beta + (1-\alpha)\gamma = P_I(coop)$$

*The possibility of a surprising result*

Now let's suppose that when the experiment is carried out, PII(coop) is found to be significantly different from PI(coop). What explanation can we give? In a classical setting, the above formula of conditional probabilities necessarily applies. If it gives a different result for the two populations, it is because the proportions involved are not the same. It is unreasonable to assume that the proportions of altruistic (respectively selfish) agents who cooperate (respectively denounce) are different in the two populations. Indeed, it is hard to see why an altruistic agent who has answered a question would play differently from an altruistic agent who has not answered any question. It must therefore be the proportions of altruistic and egoistic agents that have changed. The explanation would be that asking agents whether or not they were altruistic led some of them to change their nature: some altruists became egoists and some egoists became altruists. This explanation, although theoretically possible, nevertheless seems very strange. If all you have to do to change your preferences or character is to be asked about them, then there are doubts about the stability of these characteristics and, consequently, about the reliability of the entire model that is based on taking them into account.

---
[6] See Appendix 1 at the end of this article.

On the other hand, the explanation within the framework of the formalism we have presented is much more natural. Before they are asked the question (population I) or play the prisoner's dilemma (population II), the agents are in a superposed state of the type:

$$|\psi\rangle = \lambda_1 |A\rangle + \lambda_2 |E\rangle$$

where we denote $|A\rangle$ an altruistic state, and $|E\rangle$ a selfish state and where the coefficients $\lambda$ are the same at the outset for both populations (this is the assumption that the two populations are identical). This amounts to considering that the questionnaire is associated with an observable whose eigenvectors are $|A\rangle$ and $|E\rangle$. The dilemma decision situation is also associated with an observable whose eigenvectors are $|coop\rangle$ and $|den\rangle$. Let's assume that these two observables do not commute. In this case, there is no basis of common eigenvectors and we can write a basis change matrix in the form:

$$|A\rangle = \mu_{11} |coop\rangle + \mu_{12} |den\rangle$$
$$|E\rangle = \mu_{21} |coop\rangle + \mu_{22} |den\rangle$$

Hence :

$$|\psi\rangle = \lambda_1 |A\rangle + \lambda_2 |E\rangle = (\lambda_1 \mu_{11} + \lambda_2 \mu_{21}) |coop\rangle + (\lambda_1 \mu_{12} + \lambda_2 \mu_{22}) |den\rangle$$

The Population II agents are playing a direct game of prisoner's dilemma. So we have:

$$P_{II}(coop) = |(\lambda_1 \mu_{11} + \lambda_2 \mu_{21})|^2$$

The agents from population I start by answering the questionnaire and then play the prisoner's dilemma. So:

$$P_I(coop) = P_I(A) P_I(coop|A) + P_I(E) P_I(coop|E) = |\lambda_1|^2 |\mu_{11}|^2 + |\lambda_2|^2 |\mu_{21}|^2$$

So, in general $P_{II}(coop) \neq P_I(coop)$. This result is very similar to that obtained in the double-slit experiment. If we reason in the classical way, we think that whether or not we know which slit the photon has passed through (the analogue here of whether an agent is altruistic or egoistic) will not change its path (in this case, her response to the prisoner's dilemma). Whereas, in fact, in the case where no prior measurement is made, the two possible paths interfere on arrival on the screen (in this case, the two behaviours interfere at the moment of decision). In the case of population II, interference, materialised by cross terms of the type $(\lambda_1 \mu_{11})^* (\lambda_2 \mu_{21})$, are present, whereas these terms are destroyed by the first measure (the questionnaire) in the case of population I.

Our explanation therefore does not require us to assume that simply answering a questionnaire is enough to change a predetermined preference.

## 3.2 Framing

Kahneman and Tversky define the *framing* effect through a two-stage model of the decision-making process [1]. The first corresponds to the construction of a representation of the decision situation, the second to the choice itself. As they say: *"the real objects of evaluation are neither the objects of the real world nor the verbal descriptions of these objects; they are the mental representations we make of them"*. In order to take this point into account, we will model the process of constructing a mental representation in the same way as the choice process, by considering it as the analogue of a measurement and associating it with an observable. A framing process will therefore be defined as a set of alternative mental representations associated with the eigenvectors of the corresponding observable. Let's give an example of this type of modelling applied to an experiment carried out by Pruitt [6] and quoted by Selten [7].

*The decomposed prisoner's dilemma*

An initial population is presented with the prisoners' dilemma in the form of the following payoff table:

|   | C |   | D |   |
|---|---|---|---|---|
| C | 3 |   | 0 |   |
|   |   | 3 |   | 4 |
| D | 4 |   | 1 |   |
|   |   | 0 |   | 1 |

Player 1's payoffs are shown in rows in the top left-hand corners and player 2's payoffs are shown in columns in the bottom right-hand corners. So, if the first player plays C and the second plays D, the first will get nothing while the second will get 4. If the first player plays D as well as the second, they will both get 1.

A second population is presented with the same problem in decomposed form:

|   | For me | For him |
|---|--------|---------|
| C | 0 | 3 |
| D | 1 | 0 |

In this second case, the win is the sum of what you keep for yourself and what you get from the other player. So if the first player plays C and the second D, the first player keeps nothing for himself and gives 3 to the other, while the second player gives nothing but keeps 1 for himself. The winnings will therefore be 0 for the first player and 4 for the second.

Despite their different presentation, the two games are rigorously equivalent and, if the agents are rational, the same results should be obtained in both populations. However, Pruitt observed that much more cooperation was obtained in the second form than in the first. Selten proposes an explanation based on the limited rationality of the agents, who would not be capable of realising the equivalence between the two games.

We propose an explanation based on the fact that the state of the agents is influenced by the presentation that is made and that, even if the initial state (before they have been presented with the game) of all the agents is identical, the agents who play in case 1 are in a different state from those who play in case 2. Let's call A and B the observables associated with presentation 1 and 2 respectively. Let us call G the observable associated with the decision itself. The eigenvectors of G associated with actions C and D will be denoted $|C\rangle$ and $|D\rangle$. Suppose that presentation 1 induces a choice of mental representations[7] (associated with eigenvectors of A) between:

$|a_1\rangle$ = the game is seen as pure fun

$|a_2\rangle$ = the game is perceived as having real stakes

Similarly, suppose that presentation 2 leads to the following choice:

$|b_1\rangle$ = the game is seen as a test of generosity

$|b_2\rangle$ = lhe game is seen as a test of intelligence

The initial state of an agent can be written on either of these bases:

$$|\psi\rangle = \alpha_1|a_1\rangle + \alpha_2|a_2\rangle = \beta_1|b_1\rangle + \beta_2|b_2\rangle = \lambda|C\rangle + \mu|D\rangle$$

Whatever its initial state, an agent subjected to one or other of the presentations will see its state projected onto one or other of the eigenstates associated with that representation. Suppose that:

$$|a_i\rangle = \gamma_{1i}|C\rangle + \gamma_{2i}|D\rangle \ pour \ i = 1, 2$$
$$|b_i\rangle = \delta_{1i}|C\rangle + \delta_{2i}|D\rangle \ pour \ i = 1, 2$$

The *framing* effect is then expressed as $p_{GA}(C) \neq p_{GB}(C)$, which means that the probability of choosing action C is not the same depending on whether you were faced with situation A or situation B. Using the results from part 2, we can see that:

$$p_{GA}(C) = p_G(C) - 2\alpha_1^* \gamma_{11}^* \alpha_2 \gamma_{12}$$
$$p_{GB}(C) = p_G(C) - 2\beta_1^* \delta_{11}^* \beta_2 \delta_{12}$$

where $p_G(C)$ is the probability of choosing C in a hypothetical situation where no presentation framework would have been used and the agent's state would have remained the initial state:

$$p_G(C) = |\lambda|^2$$

The *framing* effect will therefore be visible if

$$\alpha_1^* \gamma_{11}^* \alpha_2 \gamma_{12} \neq \beta_1^* \delta_{11}^* \beta_2 \delta_{12}$$

Our explanation is therefore based on the idea that, initially in the same superposed state of C and D, two agents confronted with two different representations of the same game will see their state projected onto two new different states. The result will be a different action when they have to choose between C and D.

## 4 Conclusion

In this article, we have proposed a model based on quantum formalism. This makes it possible to recover the results of the classical Bayesian formalism when they are correct. It also seems to provide a more satisfactory explanation of certain behavioural anomalies by avoiding the ad hoc assumptions that the authors of classical models are led to accept. Our basic hypothesis is that agents' preferences (or types) can be indeterminate (i.e. superposed in the quantum sense of the term) before a decision is made. In such a case, effects similar to the interference that occurs in physics can modify the classical behaviour of the agents and explain the anomalies observed. We have proposed an explanation for the framing phenomenon. It does not contradict the classical interpretation of framing, which admits that changing the presentation frame influences the agent's state. However, the classical formalism, which assumes that the agent's state is perfectly determined initially, does not allow us to model the phenomenon of change. It is in fact unsatisfactory to assume that an agent who is initially in a well-determined state which will lead him to choose action C, will maintain this choice if he is confronted with presentation 1 and will change it if he is confronted with situation 2. The advantage of the hypothesis of indeterminacy of the initial state is that it leaves open the possibility of different choices at the very moment when the decision has to be made. We have also proposed an experiment to test our model. It is analogous to the Young's double slit experiment and clearly demonstrates the effect of probabilistic interference in the choices made by agents in two populations. The strongest argument in favour of our hypothesis concerns the robustness of preferences. It is always possible to explain, in a classical model, the different results (if any) depending on whether or not a question has been answered. It is sufficient to assume that the simple fact of having answered has led to a change in preferences. But this assumption is unsatisfactory because we are entitled to demand a certain robustness in preferences. Our model gives a more acceptable reason for these results. We now need to set up a proper experimental protocol to put the model to the test. Further work is in progress. It concerns the extension to several successive decisions and to the notion of strategy so that it can be used in game theory.

---

[7] The description of mental representations given here is of course arbitrary and makes no claim to psychological accuracy. It is given only to illustrate our point. The point we are making is above all the possibility of the existence of different mental representations.

**Ethical Statement**

The author certify that:
- The information contained in this article present the results of his research as well as an objective discussion of these results and their importance. No known fraudulent and consciously inaccurate information is presented.
- They do not submit at the same time an article representing the same results to another journal or book.
- There are no conflicts of interest that may affect the proposed publication.
- If the authors discover an important error or an inaccuracy in the present publication, they will quickly inform the editor and consider, in agreement with the person in charge, the withdrawal of the article or the publication of the information about the error.

# Appendix 1: The prisoner's dilemma

Imagine that two prisoners, Paul and Jacques, are locked in different cells with no means of communicating. They have been arrested for a common theft. The warden comes and explains to each of them that, if he confesses to having committed the theft with the other (strategy of denunciation of the other prisoner) and the other denies their participation in the theft (strategy of cooperation between the two prisoners), this will be taken into account and he will be released while the other will get 4 years in prison, but if the other also confesses, then they will both get 3 years. On the other hand, if he denies and the other confesses, it is he who will get 4 years in prison, whereas if they both deny, for lack of evidence, they will each only get one year in prison. Such a formulation is classic in game theory and can be summarised in the table of the following figure:

|  | Paul denounces | Paul cooperates |
|---|---|---|
| Jacques denounces | 3,3 | 0,4 |
| Jacques cooperates | 4,0 | 1,1 |

*Figure :* Prisoner's Dilemma Payoff Table

The table can be read by looking at the respective winnings of each player in the box corresponding to each player's move. If Jacques cooperates and Paul denounces, the payoff is 4.0, meaning that Jacques will get 4 years in prison and Paul will go free. The difficulty of the problem is that the apparent optimal strategy is unsatisfactory: it seems that whatever Jacques's behaviour, it is in Paul's interest to denounce. Indeed, if Jacques also denounces, Paul will get 3 years in prison as opposed to 4 if he had co-operated, and if Jacques co-operates, Paul will be released instead of getting 1 year in prison if he had co-operated. However, as they both know that the other can do the same thing, they will both have to turn themselves in and each get 3 years in prison, whereas if they had both cooperated, they would only have got 1 year. The problem can be generalised to an iterated dilemma in which several successive games are played and where the notion of strategy and anticipation of the other player's reaction becomes important. It can be expected that a player with an altruistic temperament will tend to cooperate, whereas a selfish player will tend to denounce.